\pdfoutput=1
\RequirePackage{ifpdf}
\ifpdf 
\documentclass[pdftex]{sigma}
\else
\documentclass{sigma}
\fi

\usepackage{ytableau}

\newcommand{\la}{{\lambda}}

\newcommand{\vp}{{\varphi}}
\newcommand{\vpd}{\varphi^\dagger}

\newcommand{\cA}{{\cal A}}

\newcommand{\cH}{{\cal H}}
\def\Z{{\Bbb Z}}   
\def\C{{\Bbb C}}   
\def\b#1{\kern-0.25pt\vbox{\hrule height 0.2pt\hbox{\vrule
width 0.2pt \kern2pt\vbox{\kern2pt \hbox{#1}\kern2pt}\kern2pt\vrule
width 0.2pt}\hrule height 0.2pt}}

\def\STrow#1{\hbox{#1}\kern-1.35pt}

\begin{document}

\ytableausetup{mathmode, boxsize=11pt}


\renewcommand{\thefootnote}{$\star$}

\renewcommand{\PaperNumber}{086}

\FirstPageHeading

\ShortArticleName{On Af\/f\/ine Fusion and the Phase Model}

\ArticleName{On Af\/f\/ine Fusion and the Phase Model\footnote{This
paper is a contribution to the Special Issue ``Superintegrability, Exact Solvability, and Special Functions''. The full collection is available at \href{http://www.emis.de/journals/SIGMA/SESSF2012.html}{http://www.emis.de/journals/SIGMA/SESSF2012.html}}}

\Author{Mark A.~WALTON}

\AuthorNameForHeading{M.A.~Walton}

\Address{Department of Physics and Astronomy, University of Lethbridge,\\ Lethbridge, Alberta, T1K 3M4, Canada}
\Email{\href{mailto:walton@uleth.ca}{walton@uleth.ca}}

\ArticleDates{Received August 01, 2012, in f\/inal form November 08, 2012; Published online November 15, 2012}

\Abstract{A brief review is given of the integrable realization of af\/f\/ine fusion discovered recently by Korf\/f and Stroppel.  They showed that the af\/f\/ine fusion of the $su(n)$ Wess--Zumino--Novikov--Witten (WZNW) conformal f\/ield theories appears in a simple integrable system known as the phase model.  The Yang--Baxter equation leads to the construction of commuting operators as Schur polynomials, with noncommuting hopping operators as arguments. The algebraic Bethe ansatz diagonalizes them, revealing a connection to the modular $S$ matrix and fusion of the $su(n)$ WZNW model.  The noncommutative Schur polynomials play roles similar to those of the primary f\/ield operators in the corresponding WZNW model. In particular, their 3-point functions are the $su(n)$ fusion multiplicities. We show here how the new phase model realization of af\/f\/ine fusion makes obvious the existence of threshold levels, and how it accommodates higher-genus fusion.}


\Classification{81T40; 81R10; 81R12; 17B37; 17B81; 05E05}

\renewcommand{\thefootnote}{\arabic{footnote}}
\setcounter{footnote}{0}

\vspace{-2mm}

\section{Introduction}

Af\/f\/ine fusion is a natural generalization of the tensor product of representations of simple Lie algebras. It is a simple truncation thereof controlled by a non-negative integer, the level. As such, it is a basic mathematical object, found in many dif\/ferent mathematical and physical contexts.
The physical context preferred by this author is provided by conformal f\/ield theory, and the so-called Wess--Zumino--Novikov--Witten (WZNW) models (see \cite{DiFrancescoP/MathieuP/SenechalD:1997}, for example). WZNW models realize at a f\/ixed non-negative integer level $k$ a non-twisted af\/f\/ine Kac--Moody algebra~$g^{(1)}$ based on a simple Lie algebra~$g$, or $g_k$ for short. Their primary f\/ields furnish representations of~$g_k$ and their operator products are governed by the corresponding af\/f\/ine fusion algebra.

Recently, Korf\/f and Stroppel \cite{KorffC/StroppelC:2010} found a much simpler physical realization of af\/f\/ine fusion, for the~$su(n)_k$ case.  The phase model \cite{BogoliubovNM/IzerginAG/KitanineNA:1998} is an integrable multi-particle model whose integrals of motion may be diagonalized by the algebraic Bethe ansatz \cite{BogoliubovNM/IzerginAG/KitanineNA:1993, NepomechieRI:1999}.  Its integrability is not only crucial to its realization of $su(n)_k$ fusion, but also explains certain properties. The integrable, or phase-model realization of af\/f\/ine fusion raises hope that a better understanding of af\/f\/ine fusion and its physical contexts will result from its study.

This paper is meant to be a gentle, non-rigorous introduction to the phase-model realization of af\/f\/ine fusion. We hope that others will share our interest in the topic and the mathematical tools involved, and perhaps help develop them further. Other reviews can be found in \cite{KorffC:2010, KorffC:2011}.

Sections \ref{section2}--\ref{section4} constitute the introductory review. Section~\ref{section5} contains some new results: thre\-shold levels (and threshold multiplicities and polynomials) and higher-genus Verlinde dimensions are both treated in the phase model there for the f\/irst time.  Section~\ref{section6} is a short conclusion.

\section{Phase model: Hilbert space and operator algebras}\label{section2}

The set of highest weights $\lambda$ of integrable highest-weight representations $L(\lambda)$ of $su(n)$ is
\begin{gather*}
 P_+ := \left\{  (\lambda_1, \lambda_2,\ldots,\lambda_{n-1}) := \sum_{a=1}^{n-1} \lambda_a  \Lambda^a\,  \vert\,  \lambda_a\in \Z_{\ge 0} \right\}  ,
\end{gather*}
where $\Lambda^a$ is the $a$-th fundamental weight. Identifying this $su(n)$ as the horizontal subalgebra of the af\/f\/ine Kac--Moody algebra $su(n)_k$ at level $k$,
\begin{gather*}
 P_+^k  :=  \left\{  \lambda  =  [ \lambda_1, \ldots,\lambda_{n-1}, \lambda_n]  :=  \sum_{a=1}^{n}  \lambda_a  \Lambda^a\,  \vert\,  \lambda_a\in \Z_{\ge 0},\, \sum_{a=1}^{n}  \lambda_a  =  k  \right\} 
\end{gather*}
 is the set of af\/f\/ine highest weights at level $k$.

The phase model has a Hilbert space $\cH$ with basis labeled by af\/f\/ine highest weights: $\vert \lambda \rangle  =  \vert \lambda_1,\ldots, \lambda_{n-1}, \lambda_n\rangle$. The Dynkin labels are interpreted as the numbers of particles at $n$ sites on a circle, corresponding to the nodes of the af\/f\/ine Dynkin diagram. If $N_a$ denotes the number operator for site~$a$, then the level $k$ is the total number of particles
\begin{gather*}
 N_a  \vert \lambda\rangle  =  \lambda_a  \vert \lambda\rangle \quad\Rightarrow\quad N \vert \lambda\rangle  =  k  \vert \lambda\rangle  ,\qquad N  :=  \sum_{a=1}^{n} N_a  .
\end{gather*}
 The basis of states is orthonormal: $\langle  \lambda \vert \mu \rangle =  \delta_{\lambda,\mu}$. Notice that this means states of dif\/ferent levels (numbers of particles) are orthogonal.

Def\/ine operators $\vpd_i$ and $\vp_i$ that create and annihilate (respectively) particles at site $i$:
\begin{gather*}
 \vpd_i \vert\lambda\rangle  =   \vert \ldots, \lambda_{i-1}, \lambda_i+1, \lambda_{i+1},\ldots \rangle,\\
  \vp_i \vert\lambda\rangle =  \begin{cases} \vert \ldots, \lambda_{i-1}, \lambda_i- 1, \lambda_{i+1},\ldots \rangle ,
 & \la_i\ge 1,\\
   0 , & \la_i  =  0  .\end{cases} 
\end{gather*}

In the phase model, these operators obey the so-called phase algebra \cite{BogoliubovNM/IzerginAG/KitanineNA:1998,KorffC/StroppelC:2010}, generated by~$\vpd_i$,~$\vp_i$ and the number operators $N_i$, for $i\in \{1,\ldots,n\}$, with relations
\begin{gather}
 [\vp_i,\vp_j]  =  [\vpd_i, \vpd_j]  =  [N_i, N_j]  =  0 , \qquad
 [N_i, \vpd_j]  =  \delta_{i,j} \vpd_i , \qquad  [N_i, \vp_j]  =  -\delta_{i,j} \vp_i  ,\nonumber\\
 N_i (1-\vpd_i\vp_i) =  0  = (1-\vpd_i\vp_i)  N_i  ,\qquad
 [\vp_i,\vpd_j]  = 0\quad  {\rm if} \ \ i\not=j, \quad {\rm but} \   \  \vp_i\vpd_i  =  1 .\label{palg}
 \end{gather}
Notice that the commutator of $\vp_i$ and $\vpd_i$ does not appear in the def\/ining relations of this algebra. That's because the phase model is the crystal limit of the $q$-boson hopping model, as made precise in~\cite{KorffC:2010}, so that a $q$-commutator reduces at $q=0$ to the last relation of~(\ref{palg}).  The operator $\pi_i := (1-\vpd_i\vp_i)$  projects to states with no particles at site $i$ (so $\pi_i^2 = \pi_i$).

As already mentioned, the level becomes the total particle number here. Therefore, to realize the fusion of a WZNW model, which has a f\/ixed level $k$, we must restrict to a f\/ixed total particle number. Hopping operators
\begin{gather}
 a_i  :=  \vpd_{i}  \vp_{i-1} ,\qquad  i\in\{1, \ldots, n\}   \label{hopop}
\end{gather} are then important\footnote{When their action is non-trivial, the operators $\vpd_{i-1}  \vp_{i}=a_i^\dagger$
hop particles in the opposite direction around the sites of the af\/f\/ine $su(n)$ Dynkin diagram. We will focus on the~$a_i$.}.
Here the indices are def\/ined mod $n$, so that $a_1 = \vpd_1 \vp_n$.\footnote{For simplicity, we put the ``magnetic f\/lux parameter'' $z$ of \cite{KorffC/StroppelC:2010} to~1.}   The action of $a_i$ is
\begin{gather}
 a_i \vert\lambda\rangle  =   \begin{cases}
 \vert \ldots,\lambda_{i-2}, \lambda_{i-1}-1, \lambda_{i}+ 1, \lambda_{i+2},\ldots \rangle  =  \vert\lambda - \Lambda^{i-1} + \Lambda^{i}\rangle  ,
 & \la_{i-1}\ge 1,\\
   0 ,  & \la_{i-1}  =  0  .
   \end{cases} \label{hop}
\end{gather}

The algebra of the hopping operators $\cA=\langle a_1,a_2,\ldots,a_n\rangle$ is def\/ined by the relations
\begin{alignat}{3}
  & \cA : \quad &&  [a_i,a_j] = 0 \qquad {\rm if}\ \ i\not = j\pm 1 \ \ {\rm mod}\ n,&\nonumber\\
  &&& a_{i}a_j^2 = a_j a_{i} a_j\quad   \&\quad a_{i}^2a_j = a_{i} a_j a_{i}\qquad  {\rm if} \ \ i=j+1 \ \ {\rm mod}\ n,    \label{alpa} &
\end{alignat}
easily verif\/ied from the phase algebra (\ref{palg}), in view of (\ref{hopop}). This algebra $\cA$ is called the af\/f\/ine local plactic algebra in \cite{KorffC/StroppelC:2010}. The f\/irst line of~(\ref{alpa}) is the locality condition. Plactic algebras were f\/irst def\/ined by Lascoux and Sch\"utzenberger, and named by them because of their relation to  tableaux (see~\cite{FultonW:1997}).

For $n=3$, the relations are
 \begin{gather}
  a_3^2a_2= a_3a_2a_3 , \qquad  a_3a_2^2= a_2a_3a_2, \qquad a_2^2a_1= a_2a_1a_2 , \qquad  a_2a_1^2= a_1a_2a_1 ,
   \nonumber\\
a_1^2a_3 =  a_1a_3a_1 , \qquad  a_1a_3^2= a_3a_1a_3.  \label{alpathree}
\end{gather}
Notice that there are no locality relations for this case~-- each node on the Dynkin diagram is a~nearest neighbour of the other~2.

When the indices of the  relations def\/ining $\cA$ are not identif\/ied mod $n$, the algebra becomes the local plactic algebra $\bar\cA=\langle a_1,a_2,\ldots,a_n\rangle$:\footnote{By abuse of notation, we use the same symbols for the generators of~$\cA$ and~$\bar\cA$. It is important to note that~$\bar\cA$  dif\/fers from the local plactic algebra ${\rm Pl_{\text{f\/in}}}$ def\/ined in \cite{KorffC/StroppelC:2010}.}
 \begin{alignat}{3}
& \bar\cA:\quad &&  [a_i,a_j] = 0\qquad {\rm if} \ \ i\not = j\pm 1, & \nonumber\\
&&&  a_{i+1}a_{i}^2 = a_i a_{i+1} a_i\quad \&\quad    a_{i+1}^2a_i = a_{i+1} a_i a_{i+1}, \qquad  1\le i\le n-1.   \label{lpa}&
\end{alignat}
This algebra is  relevant to Young tableaux and the Littlewood--Richardson algorithm that computes $su(n)$ tensor product decompositions.

The algebra $\bar\cA$ can also be realized in terms of creation operators~$\vpd_i$ and annihilation opera\-tors~$\vp_i$ obeying a phase algebra. More sites are needed, so that $i\in\{0,1,2,\ldots,n\}$, but $i=0$ and $i=n$ are not identif\/ied. Then the generators are again constructed as $a_i=\vpd_i\vp_{i-1}$, for $i\in \{1,2,\ldots,n\}$. Identifying $\vpd_0$ and $\vp_0$ with $\vpd_n$ and $\vp_n$ (respectively) then transforms the construction  for $\bar\cA$ to that for~$\cA$.

The relations def\/ining $\bar\cA$ for $n=3$ are the same as (\ref{alpathree}), except that the third line is replaced by $[a_1,a_3] = 0.$  That is,  \begin{gather}
 \bar\cA  \Rightarrow \cA: \ \ [a_1,a_n]=0 \ \mapsto\  [a_1, a_1a_n]= [a_1a_n, a_n] = 0 \label{bcAtocA}
\end{gather}
summarizes the dif\/ference between the af\/f\/ine local plactic algebra and local plactic algebra for $n=3$. Looking at~(\ref{alpa}) and~(\ref{lpa}), we see that~(\ref{bcAtocA}) applies for all~$n\ge 3$: in~$\cA$ $a_1$ and~$a_n$ do not commute, but the product~$a_1a_n$ commutes with both~$a_1$ and~$a_n$.

To see that plactic algebras are connected to Young tableaux, notice that the hopping operator $a_i$ is associated with the weight $\Lambda^{i}-\Lambda^{i-1}$. These af\/f\/ine weights have horizontal parts equal to the weights of the basic~$su(n)$ irreducible representation $L(\Lambda^1)$. The horizontal weight for~$a_{i}$ is the weight of the Young tableau
$\begin{ytableau}{i}\end{ytableau}$  that labels a vector of~$L(\Lambda^1)$~\cite{FultonW:1997}.

Now the states (vectors) of an irreducible $su(n)$ representation of highest weight $\mu$ are in 1-1 correspondence with Young tableaux of shape $\mu$ and entries in $\{1,2,\ldots,n\}$. Such a Young tableau is built starting with a Young diagram of shape $\mu$, i.e.\ one with~$\mu_1$ columns of height~1, to the right of $\mu_2$ columns of height~2, etc., up to $\mu_{n-1}$ columns of height $n-1$.  Since columns of height $n$ correspond to the trivial representation in~$su(n)$, they may be omitted.

The Young tableaux (also known as semi-standard tableaux) are obtained by f\/illing the Young diagram with entries from 1 to $n$, such that they increase going down columns, and do not decrease going across  rows~\cite{FultonW:1997}.  As an example, we display the Young tableaux for the adjoint representation of $su(3)$, of highest weight $\Lambda^1 +\Lambda^2$:
\begin{alignat}{6}
 & &&  \begin{ytableau}1 & 2\\ 2\end{ytableau}  \qquad &&  && \begin{ytableau}1 & 1\\ 2\end{ytableau} \qquad && & \nonumber\\
& \begin{ytableau}2 & 2\\ 3\end{ytableau} \qquad  && \qquad  \begin{ytableau}1 & 3\\ 2\end{ytableau} \ \ && \begin{ytableau}1 & 2\\ 3\end{ytableau}\qquad &&   &&  \begin{ytableau}1 & 1\\ 3\end{ytableau} & \nonumber\\
& &&   \begin{ytableau}2 & 3\\ 3\end{ytableau}  \qquad  &&  && \begin{ytableau}1 & 3\\ 3\end{ytableau}\qquad  && & \label{YTadj}
\end{alignat}
The arrangement is meant to remind the reader of the corresponding weight diagram. The weights of~$su(n)$ Young tableaux are determined by their entries: if there are~$\#_i$ occurrences of~$\begin{ytableau}{i}\end{ytableau}$, $i=1, \ldots, n$,  its weight is $\sum_i \#_i (\Lambda^i-\Lambda^{i-1})$.

One version of the Littlewood--Richardson rule calculates the  decomposition of the tensor-product $L(\lambda)\otimes L(\mu)$ as follows.  Take the Young tableaux of shape $\lambda$ and add them to the Young diagram of shape $\mu$, column-by-column, from right to left, to obtain ``mixed tableaux''. When adding each column, simply adjoin each box $\begin{ytableau}{i}\end{ytableau}$ to the $i$-th row of the mixed tableau. If after adding a column, the mixed tableau has an invalid shape, there is no contribution from the original Young tableau. If, on the other hand, the f\/inal mixed tableau survives, its shape $\nu$ indicates the appearance of  $L(\nu)$ in the desired decomposition.

For example, suppose $\lambda=\Lambda^1+\Lambda^2$ and $\mu=\Lambda^1$, so that the appropriate Young diagram is $\begin{ytableau}{\ }\end{ytableau}$ and the relevant Young tableaux are those of (\ref{YTadj}). Adding the rightmost column of $\begin{ytableau}1 & 3\\ 2\end{ytableau}$ to $\begin{ytableau}{\ }\end{ytableau}$ yields:
 \begin{gather*}
  \begin{ytableau}1 & 3\\ 2\end{ytableau} \ + \ \begin{ytableau}{\ }\end{ytableau} \ \Rightarrow  \  \begin{ytableau}{\ }\\ \none \\ 3\end{ytableau} 
\end{gather*}
a mixed tableau with an invalid shape, so this Young tableau produces no contribution to the tensor-product decomposition. On the other hand, we f\/ind the sequence
\begin{gather*}
 \begin{ytableau}1 & 2\\ 3  \end{ytableau} \ + \ \begin{ytableau}{\ }\end{ytableau}\ \Rightarrow  \  \begin{ytableau}{\ }\\ 2\end{ytableau}\ \ ,\ \  \begin{ytableau}{\ }& 1\\ 2\\ 3\end{ytableau} 
\end{gather*}
so the Young tableau $\begin{ytableau}1 & 2\\ 3\end{ytableau}$ reveals a representation~$L(\Lambda^1)$ in the decomposition.

Notice that adding a single box $\begin{ytableau} {i} \end{ytableau}$ to a Young diagram or mixed tableau of shape $\lambda$ produces a mixed tableaux of shape $\lambda +\Lambda^i-\Lambda^{i-1}$ or vanishes, in precise correspondence with (\ref{hop}).

As an example of the Littlewood--Richardson rule, the Young tableaux of (\ref{YTadj}) may be added to the Young diagram $\begin{ytableau} {\ }&{\ }\\ {\ }\end{ytableau}$
to verify the $su(n)$ tensor product decomposition
 \begin{gather}
L\big(\Lambda^1+\Lambda^2\big)^{\otimes 2}\ \hookrightarrow\ L(0)\oplus 2 L\big(\Lambda^1+\Lambda^2\big)\oplus L\big(3\Lambda^1\big)\oplus L\big(3\Lambda^2\big)\oplus L\big(2\Lambda^1+2\Lambda^2\big)  .
\label{tpadjadj}
\end{gather}

To understand the connection of the plactic algebra with the Littlewood--Richardson rule and Young tableaux, we must introduce words~\cite{FultonW:1997}. The (column) word of a Young tableau is obtained by listing its entries in the order from bottom to top in the left-most row, then from bottom to top in the next-to-left-most row, continuing until the top entry of the right-most row is listed. A plactic monomial is the result of substituting in the word the hopping operator $a_i$ for the number $i$.    For example, the plactic monomials of the Young tableaux in (\ref{YTadj}) are
  \begin{alignat*}{6}
  & && a_2a_1a_2 \qquad && &&  a_2a_1a_1 && &\\
&   a_3a_2a_2\qquad && \qquad  a_2a_1a_3 \ \  && a_3a_1a_2\qquad && &&  a_3a_1a_1 & \\
& && a_3a_2a_3 \qquad && &&  a_3a_1a_3\quad  && & 
\end{alignat*}
Acting on the state $\vert\mu\rangle$ with the plactic monomials of Young tableaux of a f\/ixed shape~$\lambda$ is equivalent to using the Littlewood--Richardson rule to calculate the decomposition of the tensor product~$L(\lambda)\otimes L(\mu)$.

Now, if a triple tensor product $L(\kappa)\otimes L(\lambda)\otimes L(\mu)$ is considered, the procedure is not unique. Most straightforwardly, calculating $L(\lambda)\otimes L(\mu)$ f\/irst leads to a set of mixed tableaux, one for each irreducible highest-weight representation in the decomposition. If the mixed tableaux are replaced by Young diagrams of the same shape, then the result can be calculated with the rule already described, applied a second time: $L(\kappa)\otimes (L(\lambda)\otimes L(\mu) )$.

On the other hand, one can also  multiply the Young tableaux of shape $\kappa$ and those of shape $\lambda$, to obtain a new set of Young tableaux. These product Young tableaux can then be added in the usual way to the Young diagram of shape $\mu$, to obtain the desired decomposition $(\, L(\kappa)\otimes L(\lambda)\,)\otimes L(\mu)$.  The required product $\bullet$ of Young tableaux is described by the ``bumping'' process \cite{FultonW:1997}. Fundamental examples are
\begin{gather} \ytableausetup{boxsize=normal}\begin{ytableau}
j & k
\end{ytableau}\ {\bullet\atop{}}\  \begin{ytableau} i\end{ytableau}\ {=\atop{}}\ \begin{ytableau}
i & k \\
 j
\end{ytableau}\, ,\quad i<j\le k; \qquad  \begin{ytableau}
i & k
\end{ytableau}\ {\bullet\atop{}}\  \begin{ytableau} j\end{ytableau}\ {=\atop{}}\ \begin{ytableau}
i & j \\
 k
\end{ytableau}\, ,\quad i\le j<k  . \label{fbump}
\end{gather}
 If the Young tableaux are translated into words, these bumping identities are translated into relations for the corresponding plactic algebra.  For example, with $j=k=i+1$, (\ref{fbump}) yields $a_{i+1}^2a_i = a_{i+1}a_ia_{i+1}$; and with $j=k-1=i$, we get $a_ia_{i+1}a_i = a_{i+1}a_i^2$. The full relations of (\ref{lpa}) guarantee that performing the Young tableaux calculations instead with the plactic monomials, in a fully algebraic way, will yield equivalent results.

\section[Phase model: Yang-Baxter equation and Bethe ansatz]{Phase model: Yang--Baxter equation and Bethe ansatz}\label{section3}

We now write the Yang--Baxter equation for the phase model and apply the algebraic Bethe ansatz in a standard way, following~\cite{KorffC/StroppelC:2010}. For an elementary introduction to the algebraic Bethe ansatz, see~\cite{NepomechieRI:1999}, and for a comprehensive treatment, consult~\cite{BogoliubovNM/IzerginAG/KitanineNA:1993}, for example.

First, introduce an auxiliary space isomorphic to $\C^2$, and work in $\C^2\otimes\cH$. Write
\begin{gather*}
\left(\begin{matrix}\alpha &\beta\cr \gamma & \delta\end{matrix} \right) :=
 \left(\begin{matrix}1&0\cr 0&0\end{matrix}\right)\otimes \alpha + \left(\begin{matrix}0&1\cr 0&0\end{matrix}\right)\otimes \beta + \left(\begin{matrix}0&0\cr 1&0\end{matrix}\right)\otimes \gamma + \left(\begin{matrix}0&0\cr 0&1\end{matrix}\right)\otimes \delta ,
\end{gather*}
for $\alpha$, $\beta$, $\gamma$, $\delta$ operators acting on $\cH$ (or endomorphisms of $\cH$).

Use the creation and annihilation operators of the phase model to def\/ine a Lax matrix, or $L$-operator on $\C^2\otimes\cH$
\begin{gather}
  L_i(u) :=
 \left(\begin{matrix} 1 & u\vpd_i\cr \vp_i & u\end{matrix}\right) ,\label{Liu}
 \end{gather}
where $u$ is the spectral parameter. The monodromy matrix is then
\begin{gather*}
M(u) = L_n(u) L_{n-1}(u) \cdots L_1(u) = \left( \begin{matrix}A(u) & B(u)\cr C(u) & D(u)\end{matrix} \right)  ,
\end{gather*}
 where the last equality just establishes the standard notation.
For the simple $L$-operator of (\ref{Liu}), one f\/inds $B(u) = D(u)\vpd_n$ and $C(u) = \vp_n A(u)$. For $n=3$ we f\/ind
\begin{gather}
 A(u)  =  1 + u(\vpd_2\vp_1+\vpd_3\vp_2) + u^2\vpd_3\vp_1 ,\qquad
D(u)  =  u^3 + u^2(\vp_3\vpd_2+\vp_2\vpd_1) + u\vp_3\vpd_1   .\label{ADthree}
\end{gather}

The monodromy matrix satisf\/ies the fundamental relation
\begin{gather}
R_{12}(u/v) M_1(u) M_2(v) = M_2(v) M_1(u) R_{12}(u/v) ,\label{RTT}
\end{gather}
 with $R$-matrix given by
\begin{gather*}
 R(x) = \left( \begin{matrix}\frac{x}{x-1} & 0 & 0 & 0 \cr 0 & 0 & \frac{x}{x-1} & 0 \cr 0 & \frac{1}{x-1} & 1 & 0 \cr 0 & 0 & 0 & \frac{x}{x-1} \end{matrix}\right) .    
\end{gather*}
This relation works on $\cH$ extended by two copies of the auxiliary $\C^2$, and the indices 1 and 2 indicate which of them the operators implicate.

The $R$-matrix satisf\/ies the quantum Yang--Baxter equation
\begin{gather}
 R_{12}(u/v) R_{13}(u) R_{23}(v) = R_{23}(v) R_{13}(u) R_{12}(u/v) .  \label{qYB}
\end{gather}
On the other hand, (\ref{RTT}) def\/ines the so-called quantum Yang--Baxter algebra, satisf\/ied by the entries $A(u)$, $B(u)$, $C(u)$, $D(u)$ of the monodromy matrix. For example, we f\/ind
\begin{gather}
 [B(u), B(v)] = 0,  \label{BB}
\end{gather}
important here.  The commuting $B(u)$ will be used as creation operators for a basis of states in the phase-model Hilbert space.

To demonstrate integrability, def\/ine the transfer matrix
\begin{gather}
 T(u) := \operatorname{tr}
 M(u) = A(u) + D(u),\label{transfer}
 \end{gather}
 where the trace is over the auxiliary space. Equation~(\ref{RTT}) guarantees that
 \begin{gather}
 [T(u), T(v)] =0,\label{TuTv}
\end{gather}
so that $T(u)= \sum_r u^r T_r$ is the generating function of integrals of motion: $[T_r,T_s]=0$. The Hamiltonian of the phase model is recovered as \begin{gather*}
H = -\frac1 2 ( T_1+ T_{n-1}) =  -\frac1 2 \sum_{i=1}^n\big(\vp_i\vpd_{i+1}+ \vpd_i\vp_{i+1}\big) . 
\end{gather*}

By (\ref{ADthree}), the transfer matrix is
\begin{gather*}
T(u) =  1 + u\big(\vpd_2\vp_1+\vpd_3\vp_2 + \vpd_1\vp_3\big) + u^2\big(\vpd_3\vp_1 + \vpd_2\vp_3 + \vpd_1\vp_2\big) + u^3\\
 \hphantom{T(u)}{}
 =  1 + u(a_2+a_3+a_1) + u^2(a_3a_2+a_2a_1+a_1a_3) + u^3  . 
\end{gather*}
  The general result \cite{KorffC/StroppelC:2010} is
  \begin{gather}
T(u) = \sum_{r=0}^n u^r e_r(\cA) ,  \label{Tue}
\end{gather}
where $e_r(\cA)$ indicates the $r$-th cyclic elementary symmetric polynomial, the sum of all cyclically ordered products of $r$ distinct hopping operators~$a_i$
\begin{gather}
e_r(\cA) = \sum_{\vert I\vert=r} \prod_{i\in I}^{\circlearrowleft}  a_i .\label{erA}
\end{gather}  In a monomial $a_{i_1}a_{i_2}\cdots a_{i_r}$, the relative order of~2 operators $a_{i_j}$, $a_{i_k}$ only matters if~$i_j$ and~$i_k$ dif\/fer by~1 mod\,$n$, because of~(\ref{alpa}). Suppose the nodes of the circular af\/f\/ine $su(n)$ Dynkin diagram are numbered from~1 to~$n$ in clockwise fashion. Then anticlockwise cyclic ordering $\circlearrowleft$ specif\/ies that if $i_k=i_j+1$ mod\,$n$, then $a_{i_j}$ occurs to the right of~$a_{i_k}$.  Thus, for $n=3$, $a_3a_2$ is anticlockwise circularly ordered, while $a_2a_3$ is not~-- the latter is connected to the ``long way'' anticlockwise around the circular Dynkin diagram.

By (\ref{TuTv}), we know that
\begin{gather}
 [e_r(\cA), e_{r'}(\cA) ] = 0.\label{ererp}
\end{gather}
For $n=3$,
the only non-trivial relation is
\begin{gather*}
 [ e_1(\cA), e_2(\cA) ] = [ a_1+a_2+a_3,  a_3a_2+a_2a_1+a_1a_3 ] =  0 .
\end{gather*}
Rewriting the def\/ining relations (\ref{alpathree}) as
\begin{gather*}
 [a_3, a_3a_2 ]  =  [a_3, a_1a_3]  =  [a_2, a_2 a_1]  =  [a_2, a_3a_2]  =  [a_1, a_1a_3 ]   =   [a_1, a_2a_1]  =   0
\end{gather*}
makes obvious that it is satisf\/ied.

The integrals of motion $e_r(\cA)$ are related to Schur polynomials. By the substitution $a_i\to x_i$, one recovers the elementary symmetric polynomials $e_r(x)=s_{\Lambda^r}(x)$, the Schur polynomials for the fundamental $su(n)$ representations. The def\/inition~(\ref{erA}) therefore produces noncommutative Schur polynomials for the fundamental representations of $su(n)$. The integrability result~(\ref{ererp}) allows us to def\/ine noncommutative Schur polynomials for all $su(n)$ representations. Since the~$e_r(\cA)$ commute, the Jacobi--Trudy formula
\begin{gather}
 s_\lambda(\cA)  =  \det\big( e_{\lambda_i^t-i+j}(\cA)  \big)    \label{JT}
\end{gather}
makes sense. Here $\lambda_i^t$ is the $i$-th integer of $\lambda^t$, the  transpose of the partition specifying $\lambda$.

For example, with $n=3$ and $\lambda=\Lambda^1+\Lambda^2$, we f\/ind
\begin{gather}
  s_{\Lambda^1+\Lambda^2}(\cA) = \det\left(\begin{matrix}e_2(\cA) & e_3(\cA) \cr e_0(\cA) & e_1(\cA) \end{matrix} \right)  =  \det\left(\begin{matrix}{{a_3a_2+a_2a_1+a_1a_3}} & 1 \cr 1 & {{a_1+a_2+a_3}}  \end{matrix} \right)
\nonumber\\
\hphantom{s_{\Lambda^1+\Lambda^2}(\cA)}{}
=   a_2a_1^2+a_1a_3a_1+a_3a_2^2+a_2a_1a_2+a_3a_2a_3+a_1a_3^2  \nonumber\\
\hphantom{s_{\Lambda^1+\Lambda^2}(\cA)=}{}
 + (a_3a_2a_1+a_1a_3a_2+a_2a_1a_3-1) . \label{sladj}
\end{gather} The terms of vanishing weight are enclosed in brackets.

Furthermore, the integrability (\ref{ererp}) implies that the noncommutative Schur polynomials commute among themselves\footnote{Terminology aside, it may be surprising that the noncommutative Schur polynomials commute. It was shown in~\cite{FominS/GreeneC:1998}, however, that the case studied here is but one of a more general class of such noncommutative Schur polynomials, that commute among themselves. The noncommutative arguments need only satisfy relations that are implied by those in (\ref{alpa}) but do not themselves imply~(\ref{alpa}).}
\begin{gather}
[s_\lambda(\cA), s_\mu(\cA) ] = 0. \label{slsm}
\end{gather}
One can therefore hope to f\/ind a basis diagonal in all these operators.

For that to be possible, the so-called Bethe ansatz equations must be satis\-f\/ied~\cite{BogoliubovNM/IzerginAG/KitanineNA:1993,NepomechieRI:1999}.  In more detail, the Bethe state (or vector) uses the commuting operators $B(u)$ as creation operators to construct basis elements from the vacuum
\begin{gather}
\vert b(x)\rangle := B\big(x_1^{-1}\big) B\big(x_2^{-1}\big)\cdots B\big(x_k^{-1}\big)\vert 0\rangle  ,   \label{Bsvac}
\end{gather}
 that depend on invertible indeterminates $x=(x_1,\ldots,x_k)$. Since $[N,B(u)]=B(u)$,
 each operator $B(x_i^{-1})$ injects a particle, and the states (\ref{Bsvac}) have level $k$. Recalling~(\ref{BB}), we see that $\vert b(x)\rangle$ is completely symmetric in the variables $x_1^{-1},\ldots,x_k^{-1}$.
  This can be made completely explicit using level-$k$ symmetric polynomials \cite{KorffC/StroppelC:2010}
  \begin{gather}
 \vert b(x)\rangle = \sum_{\lambda\in P_+^k} s_{\lambda^t}\big(x_1^{-1},\ldots,x_k^{-1}\big) \vert\lambda\rangle .  \label{Bvecsymm}
\end{gather}

Now the Bethe vector $\vert b(x)\rangle$ can be shown to be an eigenvector of the transfer matrix (\ref{transfer})
\begin{gather}
T(u) \vert b(x)\rangle = \left\{ \big[ 1+(-1)^ke_k(x) u^{n+k} \big] \prod_{i=1}^k \frac 1{1-ux_i} \right\} \vert b(x)\rangle ,  \label{Tubx}
\end{gather}
using the Yang--Baxter algebra (which follows from the fundamental relation (\ref{RTT})) and properties of the 0-particle vacuum $|0\rangle$ \cite{KorffC/StroppelC:2010}\footnote{In (\ref{Tubx}), $e_k(x)=x_1\cdots x_k$  is the $k$-th elementary symmetric polynomial.}.  But this works only if $x$ obeys the Bethe ansatz equations
\begin{gather}
x_1^{n+k} = \cdots = x_k^{n+k} = (-1)^{k-1} x_1 x_2  \cdots x_k .  \label{Bae}
\end{gather}

Remarkably, the solutions to (\ref{Bae}) are in 1-1 correspondence with weights in $P_+^k$. To see roughly how this works, consider the variables $y_i=x_i^{-1}x_{i+1}$, with indices def\/ined cyclically mod\,$k$.  Think of a pie that can be divided into~$n+k$ equal portions of angles $2\pi/(k+n)$ \cite{AltschulerD/BauerM/ItzyksonC:1990}. Each~$y_i$ is an $(n+k)$-th root of unity, and so determines a slice with a number of portions, the slice size. Since $y_1y_2\cdots y_k=1$, each $x=(x_1,\ldots,x_k)$ determines a slicing of the pie into~$k$ slices, or a $k$-slicing.  Furthermore, there is an $n$-slicing complementary to each $k$-slicing: where the pie is cut and where it is not cut are interchanged. The slice sizes in the $n$-slicing give the Dynkin labels of shifted weights  $\sigma+\rho$ and thus the weights $\sigma\in P_+^k$. The solutions to the Bethe ansatz equations can therefore be labelled by these $\sigma\in P_+^k$:   $x=x_\sigma$.

For complete detail, see~\cite{KorffC/StroppelC:2010}. The result, valid for all~$n$ and~$k$, is that the solutions to the Bethe ansatz equations, or Bethe roots, are in 1-1 correspondence with weights in~$P_+^k$.

\section{Af\/f\/ine fusion}\label{section4}

With the Bethe ansatz equations satisf\/ied at $x=x_\sigma$, so is equation~(\ref{Tubx}).  Then $\vert b(x_\sigma)\rangle$ is an eigenvector of the transfer matrix, and an eigenvector of all the $e_r(\cA)$, in view of~(\ref{Tue}). The eigenvalues can be determined from  (\ref{Tubx}), and one f\/inds \begin{gather*}
e_r(\cA)  \vert b(x_\sigma)\rangle = h_r(x_\sigma) \vert b(x_\sigma)\rangle , 
\end{gather*}
where $h_r(x)$ is the $r$-th complete symmetric polynomial. The noncommutative Jacobi--Trudy formula (\ref{JT}) then implies
\begin{gather}
 s_\lambda(\cA) \vert b(x_\sigma)\rangle = \det\left( h_{\lambda^t_i-i+j}(x_\sigma)\right)\vert  b(x_\sigma)\rangle  = s_{\lambda^t}(x_\sigma) \vert b(x_\sigma)\rangle . \label{slab}
\end{gather}
The last equality follows from a well-known identity for symmetric polynomials, an alternative, dual Jacobi--Trudy formula.

The connection with af\/f\/ine fusion now becomes clear, because
\begin{gather}
s_{\lambda^t}(x_\sigma) = \frac{S_{\lambda,\sigma}}{S_{{k\Lambda^n},\sigma}}.\label{slatxsi}
\end{gather}
Here $S_{\lambda,\sigma}$ denotes an element of the unitary modular $S$-matrix \cite{KacVG/PetersonDH:1984} for  $su(n)_k$. Since
\begin{gather}
{}^{(k)}N_{\lambda,\mu}^\nu = \sum_{\kappa\in P_+^k} \frac{S_{\lambda,\kappa}
S_{\mu,\kappa} S_{\nu^*,\kappa}}{S_{k\Lambda^n,\kappa}}
 \label{Vform}
\end{gather}
by the Verlinde formula \cite{VerlindeE:1988}, the  fusion eigenvalues $S_{\lambda,\sigma}/S_{k\Lambda^n,\sigma}$ obey
 \begin{gather*}
 \left(\frac{S_{\lambda,\sigma}}{S_{{k\Lambda^n},\sigma}}\right) \left(\frac{S_{\mu,\sigma}}{S_{{k\Lambda^n},\sigma}}\right) =
  \sum_{\nu\in P_+^k}  {}^{(k)}N_{\lambda,\mu}^\nu  \left(\frac{S_{\nu,\sigma}}{S_{{k\Lambda^n},\sigma}}\right) .
 \end{gather*}
 Therefore,  (\ref{slab}) and (\ref{slatxsi}) combine into
 \begin{gather*}
 s_\lambda(\cA)
 \vert b(x_\sigma)\rangle =  \frac{S_{\lambda,\sigma}}{S_{k\Lambda^n,\sigma}} \vert b(x_\sigma)\rangle ,
 \end{gather*}
 so that
 \begin{gather*}
 s_\lambda(\cA) s_\mu(\cA)  \vert b(x_\sigma)\rangle  =  \sum_{\nu\in P_+^k}  {}^{(k)}N_{\lambda,\mu}^\nu  \left(\frac{S_{\nu,\sigma}}{S_{{k\Lambda^n},\sigma}}\right)  \vert b(x_\sigma)\rangle  =  \sum_{\nu\in P_+^k}  {}^{(k)}N_{\lambda,\mu}^\nu  s_\nu(\cA)  \vert b(x_\sigma)\rangle  . 
 \end{gather*}

The fusion algebra is commutative, ${}^{(k)}N_{\lambda,\mu}^\nu= {}^{(k)}N_{\mu,\lambda}^\nu$. It is signif\/icant that the commutativity is guaranteed here by integrability: the noncommutative Schur polynomials commute by~(\ref{slsm}) because they are integrals of motion, existing due to the Yang--Baxter equation~(\ref{qYB}).

By the Bethe ansatz equations, the Bethe vectors $\vert b(x_\sigma)\rangle$ for $\sigma\in P_+^k$ form a complete orthogonal (but not normalized) basis of the Hilbert space at level $k$. Going back to~(\ref{Bvecsymm}), we can relate them to the standard basis
\begin{gather*}
\vert b(x_\sigma)\rangle  =  \sum_{\lambda\in P_+^k} s_{\lambda^t}\big(x_\sigma^{-1}\big) \vert\lambda\rangle = \sum_{\lambda\in P_+^k} \frac{S^*_{\lambda,\sigma}}{S_{k\Lambda^n,\sigma}} \vert\lambda\rangle  . 
\end{gather*}
The unitarity of the modular $S$-matrix then yields
\begin{gather*}
 \sum_{\sigma\in P_+^k} S_{k\Lambda^n,\sigma} S_{\sigma,\mu} \vert b(x_\sigma)\rangle = \vert\mu\rangle ,  
\end{gather*}
and then applying $s_\lambda(\cA)$ leads to
\begin{gather}
s_\lambda(\cA) \vert\mu\rangle = \sum_{\nu\in P_+^{k}} {}^{(k)}N_{\lambda,\mu}^\nu \vert\nu\rangle  , \label{slamuN}
\end{gather}
taking the Verlinde formula into account.

Since ${}^{(k)}N_{\lambda, k\Lambda^n}^\nu = \delta_\lambda^\nu$ (the highest weight $k\Lambda^n$ labels the identity f\/ield), we f\/ind
\begin{gather*}
s_\lambda(\cA) \vert k\Lambda^n\rangle  =   \vert\lambda\rangle  . 
\end{gather*}
This is highly reminiscent of the state-f\/ield correspondence in conformal f\/ield theory (see~\cite{DiFrancescoP/MathieuP/SenechalD:1997}), hinting that the operators $s_\lambda(\cA)$ play the role in the phase model of the primary f\/ields in the corresponding WZNW model.

This becomes clear, however, when
 \begin{gather}
\langle\nu\vert  s_\lambda(\cA) \vert\mu\rangle  =  {}^{(k)}N_{\lambda,\mu}^\nu \qquad  \forall\,\nu, \mu\in P_+^k \label{threept}
 \end{gather} and
\begin{gather*}
{}^{(k)}N_{\lambda,\mu, \nu}  = \langle k\Lambda^n\vert  s_\lambda(\cA)  s_\mu(\cA)  s_\nu(\cA) \vert k\Lambda^n \rangle   
\end{gather*}
 are written.  Indeed, the noncommutative Schur polynomials play the role of primary f\/ields, for any number of them
\begin{gather*}
{}^{(k)}N_{\lambda_1,\lambda_2,\ldots,\lambda_N}   =  \langle\lambda_1^*\vert  s_{\lambda_2}(\cA) \cdots  s_{\lambda_{N-1}}(\cA) \vert\lambda_N\rangle  =   \langle k\Lambda^n\vert  s_{\lambda_1}(\cA) \cdots  s_{\lambda_{N}}(\cA) \vert k\Lambda^n\rangle .
\end{gather*}

Like the noncommutative Schur polynomials, the Bethe vectors have an af\/f\/ine-fusion-algebraic signif\/icance \cite{KorffC:2011}. Def\/ine the (non-normalized) vectors
\begin{gather*}
\vert  b_\sigma\rangle =  \frac{\vert b(x_\sigma)\rangle}{\langle b(x_\sigma)\vert b(x_\sigma) \rangle} =
S_{k\Lambda^n,\sigma}  \sum_{\lambda\in P_+^k}  {S^*_{\lambda,\sigma}}  \vert\lambda\rangle .  
\end{gather*}
Then by Verlinde's formula (\ref{Vform}),  the formal fusion product $\ast$ yields
\begin{gather*}
  \vert\lambda\rangle \ast \vert\mu\rangle =  \sum_{\nu\in P_+^{k}} {}^{(k)}N_{\lambda,\mu}^\nu \vert\nu\rangle\quad\Rightarrow\quad \vert  b_\sigma\rangle \ast \vert  b_\tau\rangle = \delta_{\sigma,\tau} \vert  b_\sigma\rangle.  
\end{gather*}
That is, rescaled Bethe vectors  are the idempotents of the af\/f\/ine fusion algebra.

\section{New perspective}\label{section5}

Af\/f\/ine fusion appears in other integrable models~-- see \cite{GoodmanF/NakanishiT:1991}, for early examples. The simple realization af\/forded by the phase model~\cite{KorffC/StroppelC:2010}, however, provides a fresh, new perspective on the old subject.  In this section we start to exploit it.

\subsection{Threshold level}

Perhaps the most striking property of the central result~(\ref{slamuN}) is  how the level-dependence of fusion is realized. The noncommutative Schur polynomial $s_\lambda(\cA)$ has no dependence on the level! At the price of noncommutativity, the same $s_\lambda(\cA)$ works for all levels $k$. In the expression $s_\lambda(\cA)\,\vert\mu\rangle$, all level-dependence lies in the state $\vert\mu\rangle$, a much simpler object.

Af\/f\/ine fusion has a simple dependence on the level, described well by the concept of a  threshold level \cite{CumminsCJ/MathieuP/WaltonMA:1991, KirillovAN/MathieuP/SenechalD/WaltonMA:1993}.  Each highest weight representation in the decomposition of a fusion will appear at all levels greater than or equal to a minimum, non-negative integer value.  This threshold level is best understood as a consequence of the Gepner--Witten depth rule \cite{GepnerD/WittenE:1986}, or a ref\/inement thereof, conjectured in \cite{KirillovAN/MathieuP/SenechalD/WaltonMA:1993} and proved in \cite{FeingoldAJ/FredenhagenS:2008}.

All possible fusion decompositions can be given simply by treating the level as a variable, and writing multi-sets of threshold levels as subscripts.  For example, we rewrite the $su(3)$  tensor product decomposition (\ref{tpadjadj}) as
\begin{gather}
L\big(\Lambda^1+\Lambda^2\big)^{\otimes 2} \hookrightarrow  L(0)_2 \oplus   2 L\big(\Lambda^1+\Lambda^2\big)_{2, 3} \oplus   L\big(3\Lambda^1\big)_3 \oplus   L\big(3\Lambda^2\big)_3 \oplus   L\big(2\Lambda^1+2\Lambda^2\big)_4 . \label{fpadjadj}
\end{gather}
 A multi-set of threshold levels can  be replaced by a threshold polynomial $T(t)_{\lambda,\mu}^\nu$ with non-negative integer coef\/f\/icients~\cite{IrvineS/WaltonMA:2000};
 so we can also write
 \begin{gather*}
 L\big(\Lambda^1\!+\Lambda^2\big)^{\otimes 2} \hookrightarrow t^2L(0)\oplus  \big(t^2+t^3\big)L\big(\Lambda^1\!+\Lambda^2\big)\oplus  t^3L\big(3\Lambda^1\big) \oplus  t^3L\big(3\Lambda^2\big)\oplus  t^4L\big(2\Lambda^1\!+2\Lambda^2\big)  .
 \end{gather*}
 In general, the threshold polynomials are
 \begin{gather*}
 T(t)_{\lambda,\mu}^\nu = \sum_{t'}^\infty  {}^{(t')}n_{\lambda,\mu}^\nu  t^{t'}  .
 \end{gather*}
 Here the threshold multiplicities ${}^{(t)}n_{\lambda,\mu}^\nu$ satisfy
 \begin{gather*}
  {}^{(k)}N_{\lambda,\mu}^\nu =  \sum_{t}^{k}\, {}^{(t)}n_{\lambda,\mu}^\nu,  
\end{gather*} so that
$T(1)_{\lambda,\mu}^\nu   =  {}^{(\infty)}N_{\lambda,\mu}^\nu  =   T_{\lambda,\mu}^\nu$, 
the tensor-product multiplicities.
We also f\/ind
\begin{gather}
{}^{(k)}n_{\lambda,\mu}^\nu  =  {}^{(k)}N_{\lambda,\mu}^\nu  -   {}^{(k-1)}N_{\lambda,\mu}^\nu  ,\label{nNN}
\end{gather}
where we have put ${}^{(k-1)}N_{\lambda,\mu}^\nu=0$ if any of $\lambda$, $\mu$, $\nu$ are not in $P_+^{k-1}$.

In a similar way, the level-dependence can be incorporated into  (\ref{slamuN}) simply by using $\vert \mu\rangle$ with variable level. The fusion decomposition (\ref{fpadjadj}) can be derived easily this way by applying (\ref{sladj}) to $\vert \Lambda^1+\Lambda^2 +(k-2)\Lambda^3\rangle$, for example.

More generally, write $\bar\mu = \mu_1\Lambda_1+\cdots +\mu_{n-1}\Lambda^{n-1}$ and def\/ine $\bar\mu_k  :=  \bar\mu +( k-\mu_1-\mu_2-\cdots-\mu_{n-1})\Lambda^n$. Then
\begin{gather}
s_\lambda(\cA) \vert \bar\mu_k  \rangle  =  \sum_{\bar\nu\in P_+^k}  \sum_{t\le k}  {}^{(t)}n_{\lambda,\mu}^\nu  \vert  \bar\nu_k \rangle .  \label{varkmu}
\end{gather}
In the limit of large $k$, the tensor product is recovered, and (\ref{varkmu}) becomes
\begin{gather}
s_\lambda(\cA) \vert \bar\mu_\infty  \rangle  =  \sum_{\bar\nu\in P_+^k}  \sum_{t}^\infty  {}^{(t)}n_{\lambda,\mu}^\nu  \vert  \bar\nu_\infty \rangle  .    \label{varkmuinf}
\end{gather}
 Since $s_\lambda(\cA)$ does not depend on the level of $\lambda$, so that $\lambda\to \bar\lambda_\infty$ doesn't change anything, this justif\/ies our notation \begin{gather}
L(\bar\lambda) \otimes  L(\bar\mu)  \hookrightarrow  \bigoplus_{\bar\nu\in P_+}  T_{\bar\lambda,\bar\mu}^{\bar\nu}(t)  L(\bar\nu)  . \label{fprodgen}
\end{gather}

Another advantage of the phase-model realization of af\/f\/ine fusion is that, unlike in the WZNW model, the level is not f\/ixed~-- it is just the total particle number. Changes in level can be described in a simple, algebraic way by the operators~$\vpd_i$,~$\vp_i$ of the phase algebra~(\ref{palg}). In~\cite{KorffC/StroppelC:2010}, recursion relations involving fusion multiplicities at levels $k$ and $k+1$ were derived using this observation. Such relations are dif\/f\/icult to see in other ways\footnote{See what were called ``identities of the Feingold type'' in~\cite{WaltonMA:1994}, however, which relate fusion multiplicities at dif\/ferent levels.}.

Let us treat the threshold multiplicities (\ref{nNN}) in similar spirit. Notice that $\vpd_n  \vert \bar\mu_{k-1}\rangle  =  \vert \bar\mu_{k}\rangle$. So we calculate
\begin{gather*}
[s_\lambda(\cA), \vpd_n]  \vert\bar\mu_{k-1}\rangle  =  \sum_{\nu\in P_+^k} {}^{(k)}N_{\lambda, \mu}^\nu  \vert\bar \nu_k\rangle  -  \sum_{\nu\in P_+^{k-1}} {}^{(k-1)}N_{\lambda,\mu}^\nu\vp_n  \vert\bar\nu_{k-1}\rangle  .
\end{gather*}
 So the phase-model version of (\ref{nNN}) is
 \begin{gather}
 \langle \bar\nu_k\vert   [s_\lambda(\cA), \vpd_n]  \vert\bar\mu_{k-1}\rangle  =  {}^{(k)}n_{\lambda,\mu}^\nu  . \label{slvpnn}
 \end{gather}

Once a particular noncommutative Schur polynomial $s_\lambda(\cA)$ is calculated, the interesting operator $[s_\lambda(\cA), \vpd_n]$ is easy to write down, since
  $[a_i, \vpd_n]  =  \delta_{i,1}  \vpd_1  \pi_n$.  

\subsection{Higher-genus Verlinde dimensions}

As another new application of the phase-model realization of $su(n)_k$ af\/f\/ine fusion, we consider higher-genus fusion, i.e.\ higher-genus Verlinde dimensions~\cite{VerlindeE:1988}.

In the WZNW model, the fusion multiplicity ${}^{(k)}N_{\lambda,\mu}^\nu$ is also the dimension of the space of conformal blocks for the corresponding 3-point function, its Verlinde dimension.  The conformal blocks originate from correlation functions on the sphere with 3  points marked by the 3 primary f\/ields, and the fusion multiplicity can be represented graphically by a 3-legged vertex that arises in a degenerate limit of the marked sphere.

A sphere with $n$ marked points corresponds to a trivalent fusion graph with no loops.  But higher-genus Riemann surfaces can also be considered, and so fusion graphs with loops are allowed. For such higher-genus Riemann surfaces with marked points, the trivalent graph that results is not unique. The conformal bootstrap, however, ensures that the Verlinde dimension calculated from any of the graphs is the same. So, all dimensions can be built from the genus-0 3-point ones, for example. By this reasoning, one can see that the Verlinde formula extends to~\cite{VerlindeE:1988}
\begin{gather*}
 {}^{(k,g)}N_{\lambda_1, \lambda_2, \ldots, \lambda_N}  =  \sum_{\sigma\in P_+^k}  \left(  S_{k\Lambda^n,\sigma} \right)^{2(1-g)}  \left( \frac{S_{\lambda_1,\sigma}}{S_{k\Lambda^n, \sigma}}\right)\cdots  \left( \frac{S_{\lambda_N,\sigma}}{S_{k\Lambda^n, \sigma}}\right) .  
 \end{gather*}
  Here the left-hand side indicates the $su(n)_k$ Verlinde dimension for a genus-$g$ Riemann sphere with $N$ marked points.

In the phase-model realization, the argument above again applies, so that we can build all the required Verlinde dimensions from (\ref{threept}). So, for example,
\begin{gather*}
   {}^{(k,1)}N_{\lambda_1,\lambda_2}  =  \sum_{\alpha, \beta\in P_+^k} \! {}^{(k,0)}N_{\alpha^*,\beta}^{\lambda_1^*}   {}^{(k,0)}N_{\alpha, \lambda_2}^\beta
    =  \sum_{\alpha, \beta\in P_+^k}\! \langle \lambda_1^*\vert s_{\alpha^*} \vert\beta\rangle
     \langle\beta\vert s_\alpha \vert\lambda_2\rangle  =   \langle \lambda_1^*\vert \sum_{\alpha\in P_+^k}\!  s_{\alpha^*} s_\alpha  \vert\lambda_2\rangle .
\end{gather*}
 Here we have dropped the arguments from the noncommutative Schur polynomials, $\alpha^*$ indicates the weight charge-conjugate to $\alpha$, e.g., and we have used the completeness of the standard basis states.

Using this genus-1, 2-point function, the general Verlinde dimension can be constructed, with the nice result
\begin{gather}
 {}^{(k,g)}N_{\lambda_1,\dots,\lambda_N}  =    \langle \lambda_1^*\vert  \left( \sum_{\alpha\in P_+^k}  s_{\alpha^*} s_\alpha  \right)^g  s_{\lambda_2}\cdots s_{\lambda_{N-1}} \vert\lambda_N\rangle\nonumber\\
\hphantom{{}^{(k,g)}N_{\lambda_1,\dots,\lambda_N} }{}
 =  \langle k\Lambda^n\vert  \left( \sum_{\alpha\in P_+^k}  s_{\alpha^*} s_\alpha  \right)^g  s_{\lambda_1}  s_{\lambda_2}\cdots s_{\lambda_{N}} \vert k\Lambda^n\rangle  .\label{gNpt}
 \end{gather}
 Recall that all the noncommutative Schur polynomials commute. Notice that $\sum\limits_{\alpha\in P_+^k}  s_{\alpha^*} s_\alpha$ can be interpreted as a genus-generating operator, or handle-creation operator.

\section{Conclusion}\label{section6}

Let us f\/irst point out the new results obtained.
The existence of a threshold level for $su(n)_k$ af\/f\/ine fusion is made plain in the phase-model realization. The noncommutative Schur polynomials do not depend on the level; all dependence on $k$ lies in the basis vectors $\vert \lambda\rangle$. The threshold-polynomial notation (\ref{fprodgen}) was validated easily in the phase-model realization by (\ref{varkmuinf}). It was also shown in (\ref{slvpnn}) how threshold multiplicities may be calculated using, in addition to the noncommutative Schur polynomials of the hopping (af\/f\/ine local plactic) algebra, the creation operator $\vpd_n$ of the phase algebra.

The remarkable result (\ref{threept}) of \cite{KorffC/StroppelC:2010} was generalized to the elegant formula (\ref{gNpt}) for arbitrary Verlinde dimensions, at any genus $g$ and for any number $N$ of marked points.

Most of this paper is not original, however. The bulk of it was devoted to a non-rigorous summary of the integrable, phase-model realization of af\/f\/ine $su(n)$ fusion discovered recently by Korf\/f and Stroppel \cite{KorffC/StroppelC:2010} (also reviewed in \cite{KorffC:2010, KorffC:2011}). The goal was to provide a brief, easily accessible treatment in the hope of interesting others in this nice work. I believe that the Korf\/f--Stroppel integrable realization of af\/f\/ine fusion will help us understand better af\/f\/ine fusion, the WZNW models and perhaps more general rational conformal f\/ield theories.

\subsection*{Acknowledgements}
I thank Elaine Beltaos, Terry Gannon, Ali Nassar and Andrew Urichuk for discussions and/or reading the manuscript. This research was supported in part by a Discovery Grant from the
Natural Sciences and Engineering Research Council of Canada.

\pdfbookmark[1]{References}{ref}
\LastPageEnding


\begin{thebibliography}{99}
\footnotesize\itemsep=0pt

\bibitem{AltschulerD/BauerM/ItzyksonC:1990}
Altsch{\"u}ler D., Bauer M., Itzykson C., The branching rules of conformal
  embeddings, \href{http://dx.doi.org/10.1007/BF02096653}{\textit{Comm. Math. Phys.}} \textbf{132} (1990), 349--364.

\bibitem{BogoliubovNM/IzerginAG/KitanineNA:1998}
Bogoliubov N.M., Izergin A.G., Kitanine N.A., Correlation functions for a
  strongly correlated boson system, \href{http://dx.doi.org/10.1016/S0550-3213(98)00038-8}{\textit{Nuclear Phys.~B}} \textbf{516}
  (1998), 501--528, \href{http://arxiv.org/abs/solv-int/9710002}{solv-int/9710002}.

\bibitem{CumminsCJ/MathieuP/WaltonMA:1991}
Cummins C.J., Mathieu P., Walton M.A., Generating functions for {WZNW} fusion
  rules, \href{http://dx.doi.org/10.1016/0370-2693(91)91173-S}{\textit{Phys. Lett.~B}} \textbf{254} (1991), 386--390.

\bibitem{DiFrancescoP/MathieuP/SenechalD:1997}
Di~Francesco P., Mathieu P., S{\'e}n{\'e}chal D., Conformal f\/ield theory,
  \href{http://dx.doi.org/10.1007/978-1-4612-2256-9}{\textit{Graduate Texts in Contemporary Physics}}, Springer-Verlag, New York, 1997.

\bibitem{FeingoldAJ/FredenhagenS:2008}
Feingold A.J., Fredenhagen S., A new perspective on the {F}renkel--{Z}hu fusion
  rule theorem, \href{http://dx.doi.org/10.1016/j.jalgebra.2008.05.026}{\textit{J.~Algebra}} \textbf{320} (2008), 2079--2100,
  \href{http://arxiv.org/abs/0710.1620}{arXiv:0710.1620}.

\bibitem{FominS/GreeneC:1998}
Fomin S., Greene C., Noncommutative {S}chur functions and their applications,
  \href{http://dx.doi.org/10.1016/S0012-365X(98)00140-X}{\textit{Discrete Math.}} \textbf{193} (1998), 179--200.

\bibitem{FultonW:1997}
Fulton W., Young tableaux. With applications to representation theory and
  geometry, \textit{London Mathematical Society Student Texts}, Vol.~35,
  Cambridge University Press, Cambridge, 1997.

\bibitem{GepnerD/WittenE:1986}
Gepner D., Witten E., String theory on group manifolds, \href{http://dx.doi.org/10.1016/0550-3213(86)90051-9}{\textit{Nuclear
  Phys.~B}} \textbf{278} (1986), 493--549.

\bibitem{GoodmanF/NakanishiT:1991}
Goodman F.M., Nakanishi T., Fusion algebras in integrable systems in two
  dimensions, \href{http://dx.doi.org/10.1016/0370-2693(91)91563-B}{\textit{Phys. Lett.~B}} \textbf{262} (1991), 259--264.

\bibitem{IrvineS/WaltonMA:2000}
Irvine S.E., Walton M.A., Schubert calculus and threshold polynomials of af\/f\/ine
  fusion, \href{http://dx.doi.org/10.1016/S0550-3213(00)00404-1}{\textit{Nuclear Phys.~B}} \textbf{584} (2000), 795--809,
  \href{http://arxiv.org/abs/hep-th/0004055}{hep-th/0004055}.

\bibitem{KacVG/PetersonDH:1984}
Kac V.G., Peterson D.H., Inf\/inite-dimensional {L}ie algebras, theta functions
  and modular forms, \href{http://dx.doi.org/10.1016/0001-8708(84)90032-X}{\textit{Adv. Math.}} \textbf{53} (1984), 125--264.

\bibitem{KirillovAN/MathieuP/SenechalD/WaltonMA:1993}
Kirillov A.N., Mathieu P., S{\'e}n{\'e}chal D., Walton M.A., Can fusion
  coef\/f\/icients be calculated from the depth rule?, \href{http://dx.doi.org/10.1016/0550-3213(93)90087-6}{\textit{Nuclear Phys.~B}}
  \textbf{391} (1993), 651--674, \href{http://arxiv.org/abs/hep-th/9203004}{hep-th/9203004}.

\bibitem{BogoliubovNM/IzerginAG/KitanineNA:1993}
Korepin V.E., Bogoliubov N.M., Izergin A.G., Quantum inverse scattering method
  and correlation functions, \href{http://dx.doi.org/10.1017/CBO9780511628832}{\textit{Cambridge Monographs on Mathematical Physics}},
  Cambridge University Press, Cambridge, 1993.

\bibitem{KorffC:2010}
Korf\/f C., Noncommutative {S}chur polynomials and the crystal limit of the
  {$U_q\widehat{{\mathfrak {sl}}}(2)$}-vertex model, \href{http://dx.doi.org/10.1088/1751-8113/43/43/434021}{\textit{J.~Phys.~A: Math.
  Theor.}} \textbf{43} (2010), 434021, 20~pages, \href{http://arxiv.org/abs/1006.4710}{arXiv:1006.4710}.

\bibitem{KorffC:2011}
Korf\/f C., The {${\rm su}(n)$} {WZNW} fusion ring as integrable model: a new
  algorithm to compute fusion coef\/f\/icients, in Inf\/inite Analysis 2010~--
  {D}evelopments in Quantum Integrable Systems, RIMS K\^oky\^uroku Bessatsu,
  B28, Res. Inst. Math. Sci. (RIMS), Kyoto, 2011, 121--153, \href{http://arxiv.org/abs/1106.5342}{arXiv:1106.5342}.

\bibitem{KorffC/StroppelC:2010}
Korf\/f C., Stroppel C., The {$\widehat{{\mathfrak{sl}}}(n)_k$}-{WZNW} fusion
  ring: a combinatorial construction and a realisation as quotient of quantum
  cohomology, \href{http://dx.doi.org/10.1016/j.aim.2010.02.021}{\textit{Adv. Math.}} \textbf{225} (2010), 200--268,
  \href{http://arxiv.org/abs/0909.2347}{arXiv:0909.2347}.

\bibitem{NepomechieRI:1999}
Nepomechie R.I., A spin chain primer, \href{http://dx.doi.org/10.1142/S0217979299002800}{\textit{Internat.~J. Modern Phys.~B}}
  \textbf{13} (1999), 2973--2985, \href{http://arxiv.org/abs/hep-th/9810032}{hep-th/9810032}.

\bibitem{VerlindeE:1988}
Verlinde E., Fusion rules and modular transformations in 2{D} conformal f\/ield
  theory, \href{http://dx.doi.org/10.1016/0550-3213(88)90603-7}{\textit{Nuclear Phys.~B}} \textbf{300} (1988), 360--376.

\bibitem{WaltonMA:1994}
Walton M.A., Tensor products and fusion rules, \href{http://dx.doi.org/10.1139/p94-067}{\textit{Canad.~J. Phys.}}
  \textbf{72} (1994), 527--536.

\end{thebibliography}
\end{document}